\documentstyle[12pt]{article}

\begin{document}


\begin{titlepage}
\vspace{1cm}

\begin{centering}

{\Large \bf Tensor Operators in Noncommutative Quantum Mechanics }

\vspace{.5cm}
\vspace{1cm}

{\large Ricardo Amorim }

\vspace{0.5cm}

 Instituto de F\'{\i}sica, Universidade Federal
do Rio de Janeiro,\\
Caixa Postal 68528, 21945-970  Rio de Janeiro, Brazil\\[1.5ex]
\vspace{1cm}

\begin{abstract}
Some consequences of promoting the object of noncommutativity ${\mathbf \theta}^{ij}$ to an operator in Hilbert space are explored. Its canonical conjugate momentum is also introduced. Consequently,
a consistent  algebra involving the enlarged set of canonical operators is obtained, which permits us to construct theories that are dynamically invariant under the action of the rotation group.
In this framework it is also possible to give dynamics to the noncommutativity operator sector, resulting in new  features.
\end{abstract}

\end{centering}

\vspace{1cm}

\noindent PACS:  03.65.Fd, 11.10.Nx

\vfill

\noindent{\tt amorim@if.ufrj.br}
\end{titlepage}

\pagebreak


The first published work on space-time noncommutativity  appeared  in 1947 \cite{Snyder}, introduced by Snyder as an attempt to avoid 
singularities in quantum field theories; although in recent times this subject  has  been  basically related to string theory\cite{Strings}. 

Snyder introduced a five dimensional space-time with SO(4,1) as a symmetry group, with generators\footnote{ $A,B=0,1,2,3,4$; $\mu,\nu=0,1,2,3$. The parameter $a$   has dimension of length. Natural units are adopted, where $\hbar=c=1$ .} ${\mathbf M}^{AB}$ , satisfying the Lorentz algebra. Furthermore, he postulated the identification between coordinates and  generators of the $SO(4,1)$ algebra,
$\,\,
{\mathbf x}^{\mu}=a\,{\mathbf M}^{4\mu}\,\,,
$
 promoting in this way the space time coordinates to hermitian operators. The above identification implies in the commutation relation

\begin{equation}
\label{2}
[{\mathbf x}^\mu,{\mathbf x}^\nu] = i a^2{\mathbf M}^{\mu\nu}
\end{equation}

\noindent as well as in following identities:
$\,[{\mathbf M}^{\mu\nu},{\mathbf x}^\lambda]= i({\mathbf x}^{\mu}\eta^{\nu\lambda}-{\mathbf x}^{\nu}\eta^{\mu\lambda})\,$ and $\,
[{\mathbf M}^{\mu\nu},{\mathbf M}^{\alpha\beta}]= i({\mathbf M}^{\mu\beta}\eta^{\nu\alpha}-{\mathbf M}^{\mu\alpha}\eta^{\nu\beta}+{\mathbf M}^{\nu\alpha}\eta^{\mu\beta}-
{\mathbf M}^{\nu\beta}\eta^{\mu\alpha})\,,$
which are in accordance with four dimensional Lorentz invariance. 
\vskip .3cm
When open strings have their end points  on D-branes in the presence of a background constant B-field ,  effective gauge theories on a noncommutative space-time arise \cite{Hull,SW}. In these noncommutative field theories (NCFT's)\cite{NCFT},  relation  (\ref{2}) is  replaced by

\begin{equation}
\label{5}
[{\mathbf x}^\mu,{\mathbf x}^\nu] = i {\mathbf \theta}^{\mu\nu}
\end{equation} 

The point here is that the object of noncommutativity ${\mathbf \theta}^{\mu\nu}$ is usually assumed as a   constant antisymmetric matrix in NCFT's. This  violates  Lorentz symmetry, but permits to treat NCFT's as deformations of ordinary quantum field theories, replacing in brief 
ordinary products by Moyal products, and ordinary gauge interactions by the corresponding noncommutative ones. As it is well known, these theories present several
deceases as  nonunitarity, nonlocalizability, nonrenormalizability, $UV\,x\,IR$  mixing etc.  At least Lorentz invariance can be recovered by assuming that 
${\mathbf \theta}^{\mu\nu}$ is in fact a tensor operator with the same hierarchical level as the ${\mathbf x}$'s. This was done in \cite{Carlson} by using a convenient reduction of Snyder's algebra. As ${\mathbf x}^\mu$ and ${\mathbf \theta}^{\mu\nu}$  belong in this case to the same affine algebra,  the fields must be functions of the
eigenvalues of both ${\mathbf x}^\mu$ and ${\mathbf \theta}^{\mu\nu}$.  The results appearing in Ref. \cite{Carlson} are explored by some authors 
\cite{Morita}-\cite{Saxell}. Some of them prefer  to start from the beginning  by adopting the Doplicher, Fredenhagen and Roberts(DFR) algebra\cite{DFR}, which essentially assumes 
(\ref{5}) as well as the vanishing of the triple commutator among the coordinate operators.  The DFR algebra is based in  arguments coming from General Relativity and Quantum Mechanics (QM). Besides (\ref{5}) it also assumes that

\begin{equation}
\label{6}
[{\mathbf x}^\mu,{\mathbf \theta}^{\alpha\beta}] = 0
\end{equation}

The  results appearing in \cite{Carlson}-\cite{Saxell}  are  written in terms of  Weyl representations and the related Moyal products. They strongly depend on an integration over parameters related to the objects of noncommutativity,
with a weigh function $W(\theta)$. They use in this process the celebrated Seiberg-Witten\cite{SW} transformations.

A related subject  is given by noncommutative quantum mechanics (NCQM)
\cite{Durval}-\cite{Rosenbaum}. In NCQM, time is kept as a commutative parameter and space coordinates do not commute. These assumptions are reasonable in a non-relativistic theory. However most of the authors publishing in the area do not consider the objects of noncommutativity as Hilbert space operators.
Even those who consider this possibility do not introduce the corresponding conjugate momenta, which is necessary to display a complete canonical algebra and to implement rotation as a dynamical symmetry\cite{Iorio}.
These facts necessarily imply that the presented theories fail to be invariant under rotations. 
\vskip .5cm

In this work we  adapt the DFR algebra to non relativistic QM in the simplest way, but keeping consistency. 
The objects of noncommutativity are considered as  true operators and their conjugate momenta are introduced. This permits to display a complete and consistent algebra among the Hilbert space operators and to construct generalized angular momentum operators, obeying the $SO(D)$ algebra, and in a dynamical way, acting properly in all the sectors of the Hilbert space.  If this is not done, some fundamental objects usually employed in the literature, as the shifted coordinate operator( see (\ref{16})) , fail to properly transform under rotations. The symmetry is implemented not in a mere algebraic way, where the transformations are based in the indices structure of the variables, but it comes dynamically from the consistent action of an operator, as discussed in \cite{Iorio}.

We assume that space has arbitrary $D\geq2$ dimensions.  ${\mathbf x}^i$ and ${\mathbf p}_i$,  $i=1,2,...D$,  represent the position operator and its
conjugate momentum.   ${\mathbf \theta}^{ij}$ represent the noncommutativity operator, ${\mathbf \pi}_{ij}$ being its conjugate momentum. In accordance with the discussion above, it follows the algebra

\begin{equation}
\label{7}
[{\mathbf x}^i,{\mathbf p}_j] = i \delta^i_j\,\,,
\,\,\,\,\,\,\,
[{\mathbf \theta}^{ij},{\mathbf \pi}_{kl}] = i \delta^{ij}_{\,\,\,\,kl}
\end{equation} 

\noindent where $\delta^{ij}_{\,\,\,\,kl}=\delta^{i}_{k}\delta^{j}_{l}-\delta^{i}_{l}\delta^{j}_{k}$. Relation (\ref{5}) here reads as

\begin{equation}
\label{9}
[{\mathbf x}^i,{\mathbf x}^j] = i {\mathbf \theta}^{ij}
\end{equation}

\noindent and the triple commutator condition of the DFR algebra here is written as

\begin{equation}
\label{10}
[{\mathbf x}^i,{\mathbf \theta}^{jk}] = 0
\end{equation}

\noindent This implies that

\begin{equation}
\label{11}
[{\mathbf \theta}^{ij},{\mathbf \theta}^{kl}] = 0
\end{equation} 

\noindent For simplicity it is assumed that

\begin{equation}
\label{12}
[{\mathbf p}_i,{\mathbf \theta}^{jk}] = 0\,\,,\,\,\,\,\,\,
[{\mathbf p}_i,{\mathbf \pi}_{jk}] = 0
\end{equation}

The Jacobi identity
formed with the operators ${\mathbf x}^i$, ${\mathbf x}^j$ and ${\mathbf \pi}_{kl}$ leads to the nontrivial relation

\begin{equation}
\label{14}
[[{\mathbf x}^i,{\mathbf \pi}_{kl}],{\mathbf x}^j]- [[{\mathbf x}^j,{\mathbf \pi}_{kl}],{\mathbf x}^i]   =   - \delta^{ij}_{\,\,\,\,kl}
\end{equation}

\noindent The solution, unless trivial terms,  is given by

\begin{equation}
\label{15}
[{\mathbf x}^i,{\mathbf \pi}_{kl}]=-{i\over 2}\delta^{ij}_{\,\,\,\,kl}{\mathbf p}_j
\end{equation}

\noindent It is simple to verify that the whole set of commutation relations listed above is indeed consistent under all possible Jacobi identities. Expression (\ref{15}) suggests the  shifted coordinate operator\cite{Chaichan,Gamboa,Kokado,Kijanka,Calmet}

\begin{equation}
\label{16}
{\mathbf X}^i\equiv{\mathbf x}^i+{1\over 2}{\mathbf \theta}^{ij}{\mathbf p}_j
\end{equation}

\noindent that commutes with ${\mathbf \pi}_{kl}$. Actually, (\ref{16}) also commutes with ${\mathbf \theta}^{kl}$ and $ {\mathbf X}^j $, and satisfies a non trivial commutation relation with  ${\mathbf p}_i$  depending objects, which could be derived from

\begin{equation}
\label{17}
[{\mathbf X}^i,{\mathbf p}_j]=i\delta^i_j
\end{equation}

It is possible by now to introduce a continuous basis for a general  Hilbert space, with the aid of the above commutation relations. It is first necessary to find a maximal set of commuting operators. One can choose, for instance, a momentum basis, formed by the eigenvectors of ${\mathbf p},{\mathbf \pi}$. A coordinate basis formed by the eigenvectors 
of ${\mathbf X},{\mathbf \theta}$ can also be introduced, among other possibilities. We observe here that  it is in no way possible to form a basis involving more than one component of the original position operator ${\mathbf x}$, since their components do not commute. 

Just for completeness, let us display the fundamental relations involving those basis, namely  eigenvalue, orthogonality and completeness relations.

\begin{equation}
\label{18}
{\mathbf X}^i |{ X}',{ \theta}'>= {X'}^i|{ X}',{ \theta}'>
\,\,\,,\,\,\,{\mathbf\theta}^{ij} |{ X}',{ \theta}'>= {\theta'}^{ij}|{ X}',{ \theta}'>
\end{equation}

\begin{equation}
\label{19}
{\mathbf p}_i |{ p}',{ \pi}'>= {p'}_i|{ p}',{ \pi}'>
\,\,\,\,,\,\,{\mathbf\pi}_{ij} |{ p}',{ \pi}'>= {\pi'}_{ij}|{ p}',{ \pi}'>
\end{equation}

\begin{equation}
\label{20}
< { X}',{ \theta}'|{ X}",{ \theta}">= \delta^D (X'-X")\delta^{\frac{D(D-1)}{2}}({\theta}'-{\theta}")
\end{equation}

\begin{equation}
\label{21}
< { p}',{ \pi}'|{ p}",{ \pi}">= \delta^D(p'-p")\delta^{\frac{D(D-1)}{2}}({\pi}'-{\pi}")
\end{equation}

\begin{equation}
\label{22}
\int d^D X'\,d^{\frac{D(D-1)}{2}}{\theta'}  |{ X}',{ \theta}'><{ X}',{ \theta}'|= {\mathbf 1}
\end{equation}

\begin{equation}
\label{23}
\int d^D p'\,d^{\frac{D(D-1)}{2}}{\pi'}  |{ p}',{ \pi}'><{ p}',{ \pi}'|= {\mathbf 1}
\end{equation}

 \noindent Representations of the operators in those bases can be obtained in an usual way. For instance, the commutation relations (\ref{7},\ref{17})
and the eigenvalue relations above, unless trivial terms, yeld

\begin{equation}
\label{24}
< { X}',{ \theta}'|{\mathbf p}_i|{ X}",{ \theta}">= -i{\frac{\partial}{\partial X'^i}}\delta^D (X'-X")\delta^{\frac{D(D-1)}{2}}({\theta}'-{\theta}")
\end{equation}

\noindent             and

\begin{equation}
\label{25}
< { X}',{ \theta}'|{\mathbf \pi}_{ij}|{ X}",{ \theta}">= -i\delta^D (X'-X"){\frac{\partial}{\partial \theta'^{ij}}}\delta^{\frac{D(D-1)}{2}}({\theta}'-{\theta}")
\end{equation}

\noindent The transformations from one basis to the other are done by extended Fourier transforms. Related with these transformations is the "plane wave"
$< { X}',{ \theta}'|{ p}",{ \pi}">= N \exp ( i p" {X'}+  i{\pi"}{\theta'})$,
 where internal products are assumed, from now, in the pertinent expressions. For instance, $ p" {X'}+  {\pi"}{\theta'}=  p"_i {X'}^i+ {\frac{1}{2}} {\pi"}_{ij}{\theta'}^{ij}$. 
\vskip .5cm

Before discussing any dynamics, it seems interesting to study the generators of the group of rotations $SO(D)$. Not considering the spin sector,
we see that the usual angular momentum operator

\begin{equation}
\label{27}
{\mathbf l}^{ij}= {\mathbf x}^i{\mathbf p}^j-{\mathbf x}^j{\mathbf p}^i
\end{equation}

\noindent does not close in an algebra due to (\ref{9}). Actually

\begin{eqnarray}
\label{28}
[{\mathbf l}^{ij},{\mathbf l}^{kl}]&=&i\delta^{il}{\mathbf l}^{kj}-i\delta^{jl}{\mathbf l}^{ki}-i\delta^{ik}{\mathbf l}^{lj}+i\delta^{jk}{\mathbf l}^{li}\nonumber\\
&+&i{\mathbf \theta}^{il}{\mathbf p}^{k}{\mathbf p}^{j}-i{\mathbf \theta}^{jl}{\mathbf p}^{k}{\mathbf p}^{i}-i{\mathbf \theta}^{ik}{\mathbf p}^{l}
{\mathbf p}^{j}+i{\mathbf \theta}^{jk}{\mathbf p}^{l}{\mathbf p}^{i}
\end{eqnarray}

\noindent and so their components can not be  $SO(D)$ generators in this extended Hilbert space. It is not hard to see that, on the contrary, the  operator 

\begin{equation}
\label{29}
{\mathbf L}^{ij}= {\mathbf X}^i{\mathbf p}^j-{\mathbf X}^j{\mathbf p}^i
\end{equation}

\noindent closes in the $SO(D)$ algebra. However, to properly  act  in the $\theta,\pi$ sector, it has to be generalized  to the total angular momentum operator

\begin{equation}
\label{30}
{\mathbf J}^{ij}= {\mathbf L}^{ij}-{\mathbf \theta}^{il}{\mathbf \pi}_l^{\,\,j}+{\mathbf \theta}^{jl}{\mathbf \pi}_l^{\,\,i}
\end{equation}

\noindent As can be verified, not only

\begin{equation}
\label{31}
[{\mathbf J}^{ij},{\mathbf J}^{kl}]=i\delta^{il}{\mathbf J}^{kj}-i\delta^{jl}{\mathbf J}^{ki}-i\delta^{ik}{\mathbf J}^{lj}+i\delta^{jk}{\mathbf J}^{li}
\end{equation}

\noindent but ${\mathbf J}^{ij}$ generates  rotations in all of the Hilbert space sectors. Actually

\begin{eqnarray}
\label{32}
\delta {\mathbf X}^i&=&{i\over2}\epsilon_{kl}\,[ {\mathbf X}^i, {\mathbf J}^{kl}] =\epsilon^{ik}{\mathbf X}_k \nonumber\\
\delta {\mathbf p}^i&=&{i\over2}\epsilon_{kl}\,[ {\mathbf p}^i, {\mathbf J}^{kl}]=\epsilon^{ik}{\mathbf p}_k\nonumber \\
\delta {\mathbf \theta}^{ij}&=&{i\over2}\epsilon_{kl}\,[ {\mathbf \theta}^{ij}, {\mathbf J}^{kl}]=\epsilon^{ik}{\mathbf \theta}_k^{\,\,j}+
\epsilon^{jk}{\mathbf \theta}^i_{\,\,k}\nonumber\\
\delta {\mathbf \pi}^{ij}&=&{i\over2}\epsilon_{kl}\,[ {\mathbf p}^{ij}, {\mathbf J}^{kl}] =\epsilon^{ik}{\mathbf \pi}_k^{\,\,j}+
\epsilon^{jk}{\mathbf \pi}^i_{\,\,k}
\end{eqnarray}

\noindent have the expected form. The same occurs with ${\mathbf x}^i={\mathbf X}^i-{1\over 2}{\mathbf \theta}^{ij}{\mathbf p}_j$:
$\delta {\mathbf x}^i={i\over2}\epsilon_{kl}\,[ {\mathbf x}^i, {\mathbf J}^{kl}] =\epsilon^{ik}{\mathbf x}_k$.
  Observe that in the usual NCQM prescription, where the objects of noncommutativity are parameters or where the angular momentum operator has not been generalized, ${\mathbf X}$ fails to transform as a vector operator under $SO(D)$\cite{Chaichan,Gamboa,Kokado,Kijanka,Calmet}. The consistence of transformations (\ref{32}) comes from the fact that they are generated through the action of a symmetry operator and not from operations based on the index structure of those variables.

We would like to mention that in $D=2$ the operator ${\mathbf J}^{ij}$ reduces to ${\mathbf L}^{ij}$, in accordance with the fact that in this case ${\mathbf \theta}$ or ${\mathbf \pi}$ has only one independent component. In $D=3$, it is possible to represent ${\mathbf \theta}$ or $ {\mathbf \pi}$ by three vectors and both parts of the angular momentum operator have the same kind of structure, and so the same spectrum. An unexpected addition of angular momentum potentially arises, although the ${\mathbf \theta},{\mathbf \pi}$ sector can leave in a $j=0$ Hilbert subspace. 
Unitary  rotations are generated by $U(\omega)=\exp(-i\omega.{\mathbf J})$, while unitary translations, by $T(\lambda,\Xi)=\exp(-i\lambda .{\mathbf p}-i\Xi. {\mathbf \pi}) $. 
\vskip .5cm

To close this work, let us consider  the isotropic D-dimensional harmonic oscillator. There are several possibilities of rotational invariant Hamiltonians which  present the proper commutative 
limit\cite{Gamboa,Nair,Kijanka,Dadic}. A simple one is given by

\begin {equation} 
\label{34}
{\mathbf H}_0={\frac{1}{2m}}{\mathbf p}^2+{\frac{m\omega^2}{2}}{\mathbf X}^2
\end {equation}

\noindent since ${\mathbf X}^i$ commutes with ${\mathbf X}^j$, satisfies the canonical relation
(\ref{17}) and in the present formalism transforms according to (\ref{32}). 
This permits to construct   annihilation and creation operators  in the usual way: 
${\mathbf A}^i=\sqrt{{\frac{m\omega}{2}}}({\mathbf X}^i+{\frac{i{\mathbf p}^i}{m\omega}})$ and   ${\mathbf A}^{\dag i}=\sqrt{{\frac{m\omega}{2}}}({\mathbf X}^i-{\frac{i{\mathbf p}^i}{m\omega}})$. They satisfy the usual harmonic oscillator algebra, and   ${\mathbf H}_0$ can be written in terms of the sum  of D number operators      ${\mathbf N}^i={\mathbf A}^{\dag i}{\mathbf A}^i$, presenting  the same spectrum and the same degeneracies when compared with the ordinary QM case \cite{Cohen}. 
The ${\mathbf \theta},{\mathbf \pi}$ sector, however, is not contemplated with any dynamics
if ${\mathbf H}_0$ represents the total Hamiltonian. As the harmonic oscillator describes a system near an equilibrium  configuration, it seems interesting as well to add to (\ref{34}) a new term like

\begin {equation}
\label{35}
{\mathbf H}_\theta=
{\frac{1}{2\Lambda}}{\mathbf \pi}^2+{\frac{\Lambda\Omega^2}{2}}{\mathbf \theta}^2
\end {equation}

\noindent where $\Lambda$ is a parameter with dimension of $(lengh)^{-3}$ and $\Omega$ is some frequency. Both Hamiltonians can be simultaneously diagonalized, since they commute. So the total Hamiltonian eigenstates will be formed by the direct product of the Hamiltonian eigenstates of each sector. Let us consider the ${\mathbf \theta},{\mathbf \pi}$ sector.
Annihilation  and creation operators are respectively defined as $
{\mathbf A}^{ij}=\sqrt{{\frac{\Lambda\Omega}{2}}}({\mathbf \theta}^{ij}+{\frac{i{\mathbf \pi}^{ij}}{\Lambda\Omega}})$ and $
{\mathbf A}^{\dag \,ij}=\sqrt{{\frac{\Lambda\Omega}{2}}}({\mathbf \theta}^{ij}-{\frac{i{\mathbf \pi}^{ij}}{\Lambda\Omega}})$.
 They satisfy the oscillator algebra $
[{\mathbf A}^{ij},{\mathbf A}^{\dag \,kl}]=\delta^{ij,kl}$
 permitting to construct eigenstates of $H_\theta$ associated with  quantum numbers $n^{ij}\,$. As usual, the ground state is annihilated by ${\mathbf A}^{ij}$, and its  corresponding wave function ( in the ${\mathbf \theta},{\mathbf \pi}$ sector ) is

\begin {equation}
\label{38}
<{\mathbf \theta}'|n^{ij}=0,t>=({\frac{\Lambda\Omega}{\pi}})^{\frac{D(D-1)}{8}}\,
exp[-{\frac{\Lambda\Omega}{4}}{\theta'}_{ij}{\theta'}^{ij}]\,exp[-iD(D-1){\Omega\over 4} t]
\end {equation}

\noindent The wave functions for excited states are obtained through the application of the 
creation operator ${\mathbf A}^{\dag \,kl}$ on the fundamental state. However,
 we expect  that 
$\Omega$ might be so big that  only the fundamental level of this generalized oscillator could  be populated. This will generate only a shift in the oscillator spectrum, 
valuing $\Delta E={\frac{D(D-1)}{4}}\Omega$. 
This fact
seems to be trivial, but this new vacuum energy could generate unexpected behaviors.
Another point related with (\ref{38}) is that it gives  a natural way for introducing  the weight
function $W(\theta)$ which appears, in the context of NCFT's, in Refs. \cite{Carlson, Morita}. Actually, just considering the ${\mathbf \theta},{\mathbf \pi}$ sector,
the expectation value of any  function  $f({\mathbf \theta} )$ over the fundamental state is

\begin{eqnarray}
\label{39}
<f({\mathbf \theta})>&=&<n^{kl}=0,t|f({\mathbf \theta})|n^{kl}=0,t>\nonumber\\
&=&({\frac{\Lambda\Omega}{\pi}})^{{\frac{D(D-1)}{4}}}
\int \,d^{\frac{D(D-1)}{2}}{\theta'} \,\, f({\mathbf \theta}')\,exp[-{\frac{\Lambda\Omega}{2}}{\theta'}_{rs}{\theta'}^{rs}]\nonumber\\
&\equiv&\int \,d^{\frac{D(D-1)}{2}}{\theta'} \,W({\theta'})\, f({\mathbf \theta}')
\end{eqnarray}

\noindent where

\begin{equation}
\label{40}
W({\theta'})\equiv({\frac{\Lambda\Omega}{\pi}})^{{\frac{D(D-1)}{4}}}exp[-{\frac{\Lambda\Omega}{2}}{\theta'}_{rs}{\theta'}^{rs}]
\end{equation}

\noindent giving the expectation values

\begin{eqnarray}
\label{41}
<{\mathbf 1}>&=&1\nonumber\\
<{\mathbf \theta}^{ij}>&=&0\nonumber\\
{1\over2}<{\mathbf \theta}^{ij}{\mathbf \theta}_{ij}>&=&<{\mathbf \theta}^2>\nonumber\\
<{\mathbf \theta}^{ij}{\mathbf \theta}^{kl}>&=&{\frac{2}{D(D-1)}}\delta^{ij,kl}<{\mathbf \theta}^2>
\end{eqnarray}

\noindent with $<{\mathbf \theta}^2>\equiv\,{\frac{1}{2\,\Lambda\,\Omega}}\,\,$.

 This result permits to calculate expectation values of the physical coordinate operators. As one can verify,
$<{\mathbf x}^i>=<{\mathbf X}^i>=0$, but one can find non trivial noncommutativity contributions to the expectation values of other operators. For instance, it is easy to see from (\ref{41}) and (\ref{16}) that $<{\mathbf x}^2>=<{\mathbf X}^2>+
{2\over D}<{\mathbf \theta}^2><{\mathbf p}^2>$, where $<{\mathbf X}^2>$ and $<{\mathbf p}^2>$ are the usual QM results for an isotropic oscillator
in a given state. This shows that noncommutativity enlarges the root-mean-square deviation of the physical coordinate operator, as expected. 
This is an important result, which could be in principle measurable.
The inclusion of gauge interactions, the supersymmetrization and a possible relativistic generalization of this theory are under consideration and will be published elsewhere.

\vskip .5cm
\noindent {\bf Acknowledgment:} I am indebted to the UFRJ Group of Casimir Effect  for  important 
discussions. This work is partially supported  by CNPq and FUJB (Brazilian Research Agencies).

\end{document}